\documentclass[runningheads]{llncs}

\usepackage{amssymb}
\usepackage{amsmath}
\DeclareMathOperator*\argmax{arg\,max}
\def\mod{\mathbin{\text{mod}}}
\usepackage{hyperref}
\usepackage{caption}
\usepackage{subcaption}
\usepackage{paralist}
\usepackage{todonotes}
\usepackage{tikz}
\tikzset{>=latex}
\usetikzlibrary{positioning, quotes}
\usepackage{csquotes}
\usepackage[nofillcomment,ruled,vlined]{algorithm2e}

\title{Decomposing Triangulations\\into 4-Connected Components}
\author{Sabine Cornelsen and Gregor Diatzko}
\authorrunning{S.~Cornelsen and G.~Diatzko}
\institute{
    University of Konstanz\\
    \texttt{sabine.cornelsen@uni-konstanz.de, gregor.diatzko@uni-konstanz.de}
}

\begin{document}

\maketitle

\begin{abstract}
    A connected graph is 4-connected if it contains at least five vertices and
    removing any three of them does not disconnect it.
    A frequent preprocessing step in graph drawing is to decompose a plane graph
    into its 4-connected components and to determine their nesting structure.
    A linear-time algorithm for this problem was already proposed by Kant.
    However, using common graph data structures, we found the subroutine dealing
    with triangulated graphs difficult to implement in such a way that it
    actually runs in linear time.
    As a drop-in replacement, we provide a different, easy-to-implement
    linear-time algorithm that decomposes a triangulated graph into its
    4-connected components and computes the respective nesting structure.
    The algorithm is based on depth-first search.
\end{abstract}

\section{Introduction}

A connected graph is \emph{$k$-connected} if it contains at least $k+1$
vertices and removing at most $k-1$ vertices from it yields a connected graph.
A planar graph is a \emph{triangulation} or \emph{triangulated} if it is
\emph{simple}, i.e., there are no loops nor parallel edges, and each face is
bounded by a \emph{triangle}, i.e., a simple cycle of length three.
A triangulation is 4-connected if it does not contain a \emph{separating
triangle}, i.e., a triangle that does not bound a face. 4-connected triangulations are \emph{Hamiltionian}~\cite{whitney:1931}, i.e., they contain a spanning simple cycle.
Some algorithms or concepts for drawing planar graphs only work if the graph is
4-connected.
Prominent examples are rectangular duals~\cite{biedl/derka:jgaa16,kant/he:wg93},
the canonical 4-ordering~\cite{kant/he:wg93}, and compact visibility
representations~\cite{heWangZhang:tcs12}.

\begin{figure}
    \subcaptionbox{Triangulation\label{SUBFIG:introGraph}}{\includegraphics[page=1]{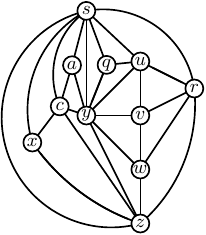}} \hfil
    \subcaptionbox{4-Block Tree\label{SUBFIG:introTree}}{\includegraphics[page=2]{figures/example}}
    \caption{\label{FIG:intro}A triangulation and the nesting structure of its 4-connected components.}
\end{figure}

In order to apply these approaches also for planar graphs that are not
4-connected, one possibility is to first triangulate the graph
\cite{biedlKantKaufmann:97} and then to split the triangulation along its
separating triangles.\footnote{For a detailed description on how to split a triangle see Page~\pageref{PARA:splitting}.} See Fig.~\ref{FIG:intro}.
After treating the \emph{4-connected components}, i.e., the thus constructed connected components (nearly)
independently, a drawing of the original graph is then constructed by combining
the drawings of the components.
See, e.g., the construction of planar L-drawings of bimodal
graphs~\cite{angelini_etal:jgaa22} or the construction of compact visibility
representations~\cite{kant:cga97}.
The \emph{4-block-tree} of a triangulation is the nesting structure of its decomposition into its 4-connected components, i.e., the vertices of the 4-block-tree are the 4-connected components and there is an edge from a triangle $t$ of a component $C$ to a component $C'$ if $t$ is the outer face of $C'$.
See Fig.~\ref{FIG:intro}.
4-block-trees are also used in order to compute
large matchings fast~\cite{rutter/wolff:acm10} and to find Tutte
paths~\cite{biedlKindermann:ICALP19} or rook-drawings with few
bends~\cite{biedlPennarun:jgaa17}.

Kanevsky et al.~\cite{kanevsky:focs91} discussed 4-connected components of
general graphs.
Kant described a method for splitting a planar graph into 4-connected components
in~\cite{kant:cga97} and gave more details for a linear time implementation in
his PhD thesis~\cite{kant:phd}.
The algorithm works as follows.
\begin{inparaenum}[(i)]
    \item
        First it splits the graph into its 3-connected components.
    \item
        Using an approach of Biedl, Kant, and
        Kaufmann~\cite{biedlKantKaufmann:97}, the separating triples are
        identified, connected by edges, and the graph is triangulated.
        If the graph was already a triangulation, the separating triangles can
        also be identified in linear time by first computing the list~$T$ of all
        triangles using the algorithm of Chiba and
        Nishizeki~\cite{chiba/nishizeki:siam_jc85} and by then removing
        those triangles from $T$ that bound a face.
    \item \label{PART:final}
        Finally the graph is split along the separating triangles and links
        between components with copies of the same triangle are established.
        In order to do so in linear time, Kant proposes to process
        the triangles in a certain order, such that for each edge~$e$ of the
        just split triangle~$t$, the interior of each separating triangle that
        contains $e$ are either all contained in the interior of $t$ or all in
        the exterior of~$t$.
        Each split is then performed in constant time.
        The latter is only possible with a special graph data structure.
        Details are described in Sect.~\ref{subsec:algoKant}.
\end{inparaenum}

When trying to implement Step~\ref{PART:final} of Kant's method, we
found it difficult to understand and to realize it in such a way that it really
works in linear time. Some problems are discussed in Sect.~\ref{subsec:problemsKant}.
In this paper we present a new and different approach for
Step~\ref{PART:final}, i.e., for computing the 4-block tree of a
triangulation.
Our algorithm is comparably easy to implement and directly yields the hierarchy
of 4-connected components of a triangulation by ordering the separating
triangles from innermost to outermost.
Splitting the triangles in this order, we can spend time proportional to the
size of the interior of the split triangle and still have an overall linear run
time. Thus, our algorithm works with any common graph data structure.
Our method is based on depth-first search and follows an approach similar to the
one used in the realization of the left-right planarity test as described by
Brandes~\cite{brandes:lr}.
An implementation of the algorithm using the OGDF framework \cite{ogdf} is
available at \url{https://gitlab.inf.uni-konstanz.de/gregor.diatzko/4connected}.

\section{Kant's Algorithm}\label{sec:algoKant}

We shortly describe a way how Kant's algorithm could be interpreted and discuss
some of its problems.

\subsection{Description}\label{subsec:algoKant}
A triangulation $G$ is \emph{split along a separating triangle} \label{PARA:splitting} $t$ with vertices $v_1,v_2,v_3$ as follows: Three new vertices $v_1',v_2',v_3'$ together with the three edges $\{v_1',v_2'\}$, $ \{v_2',v_3'\}$, and
 $\{v_3',v_1'\}$ are added to $G$. We call them a copy of $t$.
The (edges to the) neighbors of $v_i$ for $i=1,2,3$ in the interior
 of $t$ are transferred from the (incidence) adjacency list of $v_i$
to the respective list of~$v'_i$. (Alternatively the neighbors from the exterior
of $t$ can be transferred to the copy of $t$.)

In order to split the graph along the separating triangles,
Kant~\cite{kant:cga97,kant:phd} maintains the list $T$ of all separating
triangles and pointers from the vertices and edges of the separating triangles
in $T$ to the entries in the graph.
These pointers would have to be updated after a split if the resulting vertex or
edge was now a copy.
In order to avoid this, Kant suggests to split the separating triangles in a
specific order.
The idea is that whenever the graph is split at a separating triangle $t$ into
two graphs $G_1$ and $G_2$, then the remaining separating triangles that share
an edge $e$ with $t$ are either all contained in $G_1$ or all in $G_2$.
This will be the subgraph where the original edge $e$ of $t$ will stay, while the copy goes to the other subgraph.

  \begin{figure}
    \centering
    \subcaptionbox{\label{SUBFIG:pathD}}{\begin{tikzpicture}[scale=.88]
    \tikzset{every path/.style={thick}}
    \tikzset{every node/.style={circle,inner sep=0}}
    \node (v) {$v$};
    \node (dummy) [above of = v] {};
    \node (w) [above of = dummy] {$w_3$}
    edge (v);
    \node (u1) [right of = dummy] {$w_4$}
    edge (v) edge (w);
    \node (u2) [right of = u1] {$w_5$}
    edge (v) edge (w);
    \node (u3) [right of = u2] {$w_6$}
    edge (v) edge (w);
    \node (u4) [right of = u3] {$w_7$}
    edge (v) edge (w);
    \node (u8) [left of = dummy] {$w_2$}
    edge (v) edge (w);
    \node (u7) [left of = u8] {$w_1$}
    edge (v) edge (w);
    \node (u6) [left of = u7] {$w_0$}
    edge (v) edge (w);
    \node (u5) [left of = u6] {$w_8$}
    edge (v) edge (w);
    \end{tikzpicture}} \hfil
    \subcaptionbox{\label{SUBFIG:cyclicD}}{\begin{tikzpicture}[scale=.6]
    \tikzset{every path/.style={thick}}
    \tikzset{every node/.style={circle,inner sep=0}}
    \node (v1) at (0,0) {$w_2$};
    \node (v2) at (4,0) {$w_1$}
    edge (v1);
    \node (v4) at (2,1.15) {$v$}
    edge (v1) edge (v2);
    \node (v3) at (2,3.46) {$w_0$}
    edge  (v1) edge (v2) edge (v4);
    \node (u1) at (2.67,1.54) {$\bullet$}
    edge (v2) edge (v3) edge (v4);
    \node (u2) at (1.33,1.65) {$\bullet$}
    edge (v1) edge (v3) edge (v4);
    \node (u3) at (2,0.38) {$\bullet$}
    edge (v2) edge (v1) edge (v4);
    \end{tikzpicture}}
    \caption{\label{FIG:kantsorder}
        (a) Omitting transitive edges, the separating triangles sharing the edge $\{v,w_3\}$ are linked
        $\langle v,w_3,w_5\rangle \rightarrow \langle v,w_3,w_6\rangle \rightarrow \langle v,w_3,w_7\rangle \rightarrow \langle v,w_3,w_8\rangle \rightarrow \langle v,w_3,w_0\rangle \rightarrow \langle v,w_3,w_1\rangle$
         in Kant's  Algorithm. \,(b) Processing $v$ first, graph $D$ is the directed triangle
         $\langle v,w_0,w_1\rangle \rightarrow \langle v,w_0,w_2\rangle =
         \langle v,w_2,w_0\rangle \rightarrow \langle v,w_2,w_1\rangle=
         \langle v,w_1,w_2\rangle \rightarrow \langle v,w_1,w_0\rangle$.}
 \end{figure}
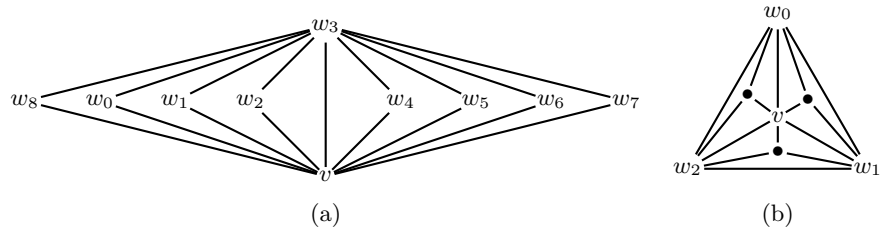

To this end Kant defines a (labeled) directed graph $D$ on the separating triangles.
The details in \cite{kant:cga97} and \cite{kant:phd} differ slightly and are
both not completely correct.
Here we give an interpretation of the version in \cite{kant:phd}:
Each separating triangle $t$ is stored at each of its three edges.
Now the vertices are processed in an arbitrary order.
For each vertex $v$, let $w_0,\dots,w_{d-1}$ be the adjacent vertices of $v$ in clockwise order around $v$.
Let $w_i$, $i=1,\dots,d-1$ be a neighbor of $v$ that was not processed before $v$.
Let $t=\langle v,w_i,w_j \rangle$ and let $t'=\langle v,w_i,w_k \rangle$ be two triangles that share the edge $\{v,w_i\}$.
Then there is an edge from $t$ to $t'$ labeled $\{v,w_i\}$ if and only if $(j-i) \mod d < (k-i) \mod d$. See Fig.~\ref{FIG:kantsorder} for examples. Observe that $D$ does not have to be acyclic. See Fig.~\ref{SUBFIG:cyclicD}. 

Now a triangle $t$ may be split if for each edge $e$ contained in $t$, the directed graph $D$ either contains no incoming edge labeled $e$ or no outgoing edge labeled $e$. Observe that an innermost separating triangle always fulfills this property. E.g., in the example in Fig.~\ref{SUBFIG:pathD}, the triangles could be split in the order $\langle v,w_3,w_i\rangle $, $i=5,6,7,8,0,1$, where the first three triangles leave the remaining separating triangles in the exterior while the last three triangles leave the remaining separating triangles in their interior.

\subsection{Problems}\label{subsec:problemsKant}

Kant's algorithm has several problems.
\begin{enumerate}
\item
The number of edges in the directed graph $D$ might be quadratic in the number of vertices of the input graph. So $D$ must not be computed entirely.
It would suffice, however, to compute a transitive reduction of $D$.
This could be done in linear time by sorting the separating triangles $\langle v,w_i,w_j \rangle$  containing the edge $\{v,w_i\}$ according to  $j-i$ using bucket sort.
\item With this approach it is impossible to maintain the pointers from the separating triangles to their vertices in the graph without updating them after a split as mentioned by Kant. However, it suffices to maintain pointers from the separating triangles to the entries of their edges in the incidence lists.
\item Since the triangles are not necessarily processed inside out, the nesting
structure is not immediately obtained during the construction.
Moreover, in each splitting step it has to be decided to which component the
copy of the separating triangle has to go; more precisely this decision has to
be made for every single edge of the separating triangle independently.
\item\label{ITEM:constantSplit}
Finally, the running time of Kant's algorithm relies on the fact that a split
can be performed in constant time.
This is true if only the adjacency lists of the vertices of the separating
triangles have to be split.
But if we also want to make sure that the neighbors of a split triangle
\enquote{know} whether they are adjacent to the original vertex or the copy,
then this is no longer true.
Moreover, if incidence lists are used instead of adjacency lists, then it is
impossible to maintain the end vertices for all edges in constant time per~%
split.
\end{enumerate}
Problem~\ref{ITEM:constantSplit} can be mitigated by maintaining incidence lists
with \enquote{symbolic} edges that do not know their end vertices.
I.e., each vertex is associated with a cyclic list.
The size of the list is unknown.
The entries of the lists are distinct identifiers.
Two vertices are adjacent if their cyclic lists contain the same identifiers.
But the cyclic list of a vertex alone contains no information about its
neighbors.
This information can only be obtained in a postprocessing step.
However, this rules out the use of any common graph data structure.
In effect, e.g., the function in OGDF for splitting vertices has a running time
that is linear in the number of edges being transferred to the copy.

To summarize, Kant's approach seems to be realizable in linear time.
However, it has several disadvantages.
Thus, we describe our new approach in the next two sections.
It is based on ordering the separating triangles from innermost to outermost.

\section{Algorithm Overview}\label{sec:algo}

Throughout the remainder of this paper, let $G=(V,E)$ be a triangulation with a
fixed outer face. We compute the 4-block tree of $G$ as follows.

\paragraph{1.~Listing separating triangles.}
First we compute the list of all triangles of~$G$ using the algorithm of Chiba
and Nishizeki~\cite{chiba/nishizeki:siam_jc85}.
Then we remove the face boundaries from that list, i.e., the triangles $\langle u,v,w \rangle$ where the edges to $v$ and $w$ appear consecutively in the
incidence list of~$u$.
This yields the list~$T$ of separating triangles of~$G$.

\paragraph{2.~Ordering separating triangles.}
In the next step, we order the separating triangles in $T$ such that if 
triangle~$t$ contains the interior of triangle~$t'$ in its interior then $t'$ is before~$t$.
E.g., in Fig.~\ref{FIG:intro}, $t_1,t_2,t_3,t_4$ would be an appropriate
ordering of the separating triangles.
In this step we also compute for each separating triangle~$t$ an oriented
reference edge~$e$ such that the interior of $t$ is to the left of~$e$.
This step is the main difference to Kant's approach~\cite{kant:cga97}.
It is based on depth-first search and described in detail in the next section.

\paragraph{3.~Splitting along separating triangles.}
We process the separating triangles in the order computed in Step~2, i.e., 
innermost triangles are considered first.
For each separating triangle~$t$ make copies $v_1',v_2',v_3'$ of its three
vertices $v_1,v_2,v_3$, and add the three edges $\{v_1',v_2'\}, \{v_2',v_3'\},
\{v_3',v_1'\}$.
Starting from the reference edge~$e$ of~$t$, transfer the edges in the interior
of $t$ that are incident to $v_i$, $i=1,2,3$ from the incidence list of $v_i$
to the incidence list of~$v'_i$:

More precisely, assume that $e=(v_1,v_2)$ is the reference edge.
Starting at~$e$ walk through the incidence list of $v_1$ in counter-clockwise
direction until $\{v_1,v_3\}$ is reached.
Remove each edge of the incidence list of $v_1$ between $\{v_1,v_2\}$ and
$\{v_1,v_3\}$ from the incidence list of $v_1$ and insert it into the incidence
list of $v'_1$.
Consider now $(v_3,v_1)$ and later $(v_2,v_3)$ as the new reference edge, and
continue.
This can be done in time linear in the number of transferred edges.
Observe that each edge is transferred at most once during the course of the
algorithm.

Now the connected component $H$ containing $v_1'$ is a 4-connected component of
$G$.
To store the nesting structure,
we set a pointer from the reference edge of $t$ to $H$.
This is a \emph{preliminary child pointer}.
E.g., assume that the reference edge of triangle~$t_1$ in Fig.~\ref{FIG:intro}
is $(s,z)$.
When we split the triangle~$t_3$, then $(s,z)$ is still an edge of (the
remainder of)~$G$.
But the triangle $t_1$ that should be the parent of $H_1$ is now a face of the
just constructed component~$H_3$.
So when we split a triangle obtaining a 4-connected component~$H$, we test
whether an involved edge~$e$ of $G$ is a reference edge of a triangle~$t'$ of
$G$ with preliminary child component~$H'$ and if so, we set the \emph{parent} of
$H'$ to be the copy~$e'$ of $e$ in~$H$.

\section{Ordering Separating Triangles from Inner- to Outermost}
\label{sec:ordering}

\begin{figure}[t]
    \centering
    \subcaptionbox{DFS tree (blue), back edges (gray), traversal order of the
    second DFS (edge labels).\label{FIG:DFS}}[\linewidth]
    {\includegraphics[page=4]{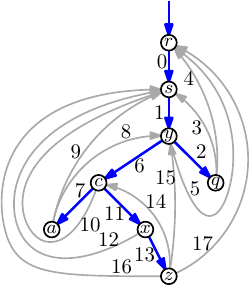}}
    \hfil
    \subcaptionbox{Ordering of the separating triangles as computed by the algorithm.\label{FIG:ordering}}[\linewidth]%
    {\includegraphics[page=8]{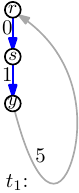} \hfil
        \includegraphics[page=7]{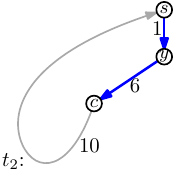} \hfil
        \includegraphics[page=6]{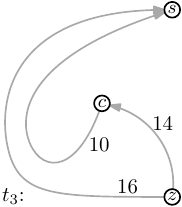} \hfil
        \includegraphics[page=5]{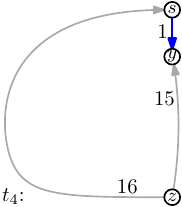}}
    \caption{\label{FIG:illustrationALGO}The maximum edge label of a triangle
    $t$ is the moment \textsc{time}[$t$] when $t$ was discovered.
    \textsc{time}[$t_1$] = 5, \textsc{time}[$t_2$] = 10, and
    \textsc{time}[$t_3$] = \textsc{time}[$t_4$] = 16. The triangles are ordered
    ascending according to \textsc{time} where ties are broken according to the
    internal angle at the head of the discovering edge. E.g., $t_3$ is before
    $t_4$ since $\| \angle zsc \| = 2 < 4 = \| \angle zsy \|$.}
\end{figure}

In this section, we show how the list of separating triangles of a triangulation can be ordered in linear
time such that innermost triangles are first, i.e., such that a triangle $t$ is before a triangle $t'$ if $t$ is contained in $t'$.
We will first describe the algorithm, then provide a proof of its correctness,
and finally give more details on how to implement it in linear time.
A complete pseudocode is provided in the appendix.

We start with a definition.
For a vertex $v$ and two of its neighbors $u$ and $w$ we denote by $\|\angle
uvw\| \in \{0, \dots,\textsc{deg}(v)\}$ the \emph{size of the angle} between the
edges $\{v,u\}$ and $\{v,w\}$ at $v$ in the following sense:
starting from the edge~$\{v,u\}$, count the number of edges one has to turn
counter-clockwise (CCW) until one sees the edge~$\{v,w\}$.
In Fig.~\ref{FIG:DFS}, we have $\|\angle qsz\| = 2$ while $\|\angle
zsq\| = 5$.
During the algorithm we want to compute the size of an angle in constant time.
To this end, before the algorithm, we number the entries in the incidence list
of each vertex $v$ in CCW order starting at an arbitrary entry;
let these numbers be $\textsc{index}_v(x)$ for the entry of the neighbor $x$ of
$v$.
Now the size of the angle $\|\angle uvw\|$ is $(\textsc{index}_v(w) -
\textsc{index}_v(u)) \mod \textsc{deg}(v)$.

Motivated by \cite{brandes:lr}, we perform depth-first search (DFS) twice,
starting from a vertex $r$ incident to the outer face.
DFS explores the graph as far as possible before backtracking using a stack.
In the following we consider edges to be directed as traversed during the first
DFS.
DFS partitions the edges of the graph into \emph{tree edges}, along which new
vertices were discovered, and \emph{back edges}, which point from a vertex to
one of its ancestors;
each edge fits into one of these two classes.
Let the tree edge along which a vertex $v \not= r$ was discovered be the
\emph{parent edge} of $v$.
For ease of exposition, we consider the parent edge of $r$ to be a virtual edge
pointing into the outer face.
For each back edge $b$, we call the unique directed cycle $\langle e_1, \dots,
e_k, b \rangle$ consisting of $b$ and tree edges $e_1, \dots, e_k$ the
\emph{fundamental cycle}~$C(b)$ of $b$.
During the first DFS, we collect the following additional information.
The \emph{depth} $\textsc{depth}(v)$ of a vertex $v$ is the length of the unique
$r$-$v$-path along tree edges.
The \emph{lowpoint} of a back edge $b = (v,w)$ is $\textsc{lowpt}(b) =
\textsc{depth}(w)$.
Let $(p,w)$ be the parent edge of $w$ and let $(w,c)$ be the first edge on
$C(b)$;
the \emph{angle} of $b$ is $\textsc{angle}(b) = \|\angle vwc\|$ if the edges to $p, v, c$
appear in this order in the CCW incidence list of $w$, otherwise it is
$\|\angle cwv\|$.
In the former case we call $b$ a \emph{left} back edge (i.e., $C(b)$ is a clockwise
cycle), in the latter a \emph{right} back edge.
An \emph{outermost return edge} of an edge $e$ is a back edge $b$ for which
$e \in C(b)$ and that maximizes $(-\textsc{lowpt}(b),\textsc{angle}(b))$~--~sorted lexicographically, i.e., a back edge with maximum angle among all back edges $b$ with minimum lowpoint among those fulfilling $e \in C(b)$.
A tree edge
$e$~inherits its lowpoint and angle from its outermost return edges.
This concludes the first DFS.
\looseness=-1

In Fig.~\ref{FIG:DFS}, for example,
$(z,y)$ is a right back edge because $s, c, z$ is the CCW order of the
three neighbors at $y$.
We have
$\textsc{parentEdge}(c) = (y,c)$,
$\textsc{lowpt}(c,x) = \textsc{depth}(r) = 0$, and
$\textsc{angle}(c,x) = \|\angle srz\| = 3$.
Thus $(z,r)$ is the only outermost return edge of the tree edge $(c,x)$
and further
$\textsc{lowpt}(c,a) = 1$ and
$\textsc{angle}(c,a) = 1$ while
$\textsc{lowpt}(c,s) = 1$ and
$\textsc{angle}(c,s) = 2$.

Before the second DFS, we sort the outgoing edges $e$ of each vertex lexicographically by
$(-\textsc{lowpt}(e), \textsc{angle}(e))$.
E.g., among the outgoing edges of $b$ in Fig.~\ref{FIG:DFS}, $(c,a)$ is before
$(c,s)$ due to its smaller angle and $(c,s)$ is before $(c,x)$ due to its
greater lowpoint.
Now we perform the second DFS following this order of edges. Observe that this still yields the same orientation and partition of the edges into tree and back edges as in the first DFS; only the order in which the subtrees are traversed may be different.
During this second traversal of the graph, we take note of the order in which
the separating triangles are discovered.
For this purpose we consider a separating triangle to be discovered once all
three of its edges have been traversed.
If several separating triangles $t_1, \dots, t_k$ are discovered at the same
time, they share an edge $(v,w)$ traversed from $v$ to $w$ and have distinct third vertices $u_1, \dots,
u_k$. Moreover, the shared edge $(v,w)$ is the edge of $t_1, \dots, t_k$ that was discovered last.
Let the \emph{internal angle} of $t_i$ be the angle at $w$ inside $t_i$; its
size is $\|\angle vwu_i\|$ if $(v,w)$ is a left back edge, and $\|\angle
u_iwv\|$ otherwise.
Now we order $t_1, \dots, t_k$ by the size of their internal angle in increasing order.
We now have an ordering of all separating triangles.
We set the \emph{reference edge} of triangle $t_i$ to be $(v,w)$ if the third
traversed edge $(v,w)$ of $t_i$ is a right back edge and $(w,v)$ otherwise.
In Fig.~\ref{FIG:DFS}, $(z,s)$ is the last edge of the separating
triangle $t= \langle z,b,s \rangle$ that is traversed. Since $(z,s)$ is a left back edge it follows that the reference edge of $t$ is $(s,z)$.
The size of the internal angle of $t$ is $\|\angle zsc\| = 2$.

\begin{theorem}
    \label{thm:correctness}
    The algorithm described above orders the separating triangles of $G$
    correctly from innermost to outermost in linear time.
\end{theorem}
\begin{proof}
Consider the point in time during the second DFS when we traverse the last edge
$(v,w)$ of a separating triangle $t = \langle u,v,w \rangle$.

\paragraph{The vertices $u$, $v$ and $w$ lie on a single directed tree path $P$
from $w$ to either $u$ or $v$:}
Since $(v,w)$ closes a cycle, it must be a back edge, implying
$\textsc{depth}(w) < \textsc{depth}(v)$.
If $\{u,v\}$ is a tree edge, then $|\textsc{depth}(u) - \textsc{depth}(v)| =
1$, and $u~\neq~w$ implies $\textsc{depth}(w) < \textsc{depth}(u)$.
If $\{u,v\}$ is the back edge $(u,v)$, we have $\textsc{depth}(w) <
\textsc{depth}(v) < \textsc{depth}(u)$.
And if $\{u,v\}$ is the back edge $(v,u)$, we know that
$\textsc{lowpt}(v,u) > \textsc{lowpt}(v,w)$ because it was traversed earlier.
Thus $\textsc{depth}(w) < \textsc{depth}(u)$ follows in all three cases, which
means that $u$ and $v$ are contained in the subtree rooted at $w$.
Since $u$ and $v$ are also connected by an edge, it follows that there is a
directed tree path $P$ from $w$ to one of $u$ or $v$, whichever is deeper, such
that all three vertices of $t$ lie on $P$.

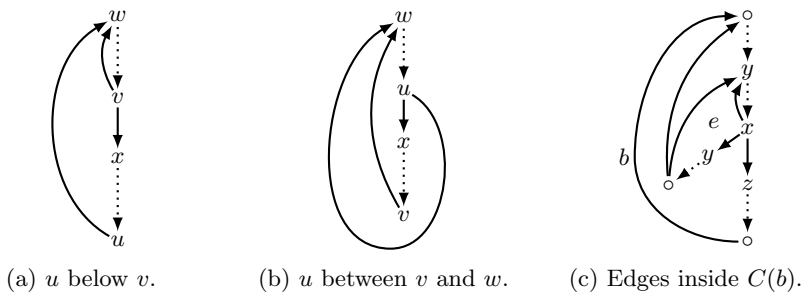
\begin{figure}[t]
    \centering
    \subcaptionbox{$u$ below $v$.\label{fig:proof:below}}[0.3\linewidth]{\begin{tikzpicture}
\tikzset{every path/.style={thick}}
\tikzset{every node/.style={circle,inner sep=0}}
\node (w) at (5,3) {$w$};
\node (v) at (5,1.9) {$v$};
\node (x) at (5,1.1) {$x$};
\node (u) at (5,0) {$u$};
\draw[->,dotted] (w) -- (v);
\draw[->] (v) -- (x);
\draw[->,dotted] (x) -- (u);
\draw[->] (v) to[bend left] (w);
\draw[->] (u) to[bend left=60] (w);
\end{tikzpicture}
}
    \hfil
    \subcaptionbox{$u$ between $v$ and $w$.\label{fig:proof:between}}[0.3\linewidth]{\begin{tikzpicture}[scale=.88]
\tikzset{every path/.style={thick}}
\tikzset{every node/.style={circle,inner sep=0}}
\node (w) at (5,4) {$w$};
\node (u) at (5,2.9) {$u$};
\node (x) at (5,2.1) {$x$};
\node (v) at (5,1) {$v$};
\draw[->,dotted] (w) -- (u);
\draw[->] (u) -- (x);
\draw[->,dotted] (x) -- (v);
\draw[->] (v) to[bend left] (w);
\draw[->] (u) to[out=330,in=0] (4.8,.5) to[out=180,in=210] (w);
\end{tikzpicture}
}
    \hfil
    \subcaptionbox{Edges inside $C(b)$.\label{fig:proof:inside}}[0.3\linewidth]{\begin{tikzpicture}[scale=.75]
\tikzset{every node/.style={circle,inner sep=0}}
\tikzset{every path/.style={thick}}
\node (w) at (5,5) {$\circ$}; 
\node (z) at (5,4) {$y$};
\node (y) at (5,3) {$x$};
\node (s) at (5,2) {$z$};
\node (v) at (5,1) {$\circ$}; 
\node (x) at (4.3,2.5) {$y$};
\node (o) at (3.6,2) {$\circ$};
\draw[->,dotted] (w) -- (z);
\draw[->,dotted] (z) -- (y);
\draw[->] (y) -- (s);
\draw[->,dotted] (s) -- (v);
\draw[->] (y) -- (x);
\draw[->,dotted] (x) -- (o);
\draw[->] (y) to[bend left] (z);
\draw[->] (o) to[bend left] (w);
\draw[->] (o) to[bend left] (z);
\draw[->] (v) to[out=180,in=270] (3,2.5) to[out=90,in=180] (w);
\node[left] (b) at (3,2.5) {$b$};
\node (e) at (4.4,3.1) {$e$};
\end{tikzpicture}
}
    \caption{Illustrations for the proof of Theorem~\ref{thm:correctness}.
        Dotted lines represent paths of length $\geq 0$.
        In (a) and (b), the edge $(v,w)$ is the last edge of triangle $t =
        \langle u,v,w\rangle$ we traverse.}
    \label{fig:proof}
\end{figure}

\paragraph{All separating triangles that are discovered with the edge
$(v,w)$ lie to the right (left, resp.) of $(v,w)$ if $(v,w)$ is a left (right,
resp.) back edge:}
(This implies that our calculation of the sizes of the internal angles as well
as the reference edge are correct.)
W.l.o.g.~assume for contradiction that $(v,w)$ is a left back edge and $t$ lies
to the left of $(v,w)$.
We have already shown that $\textsc{depth}(w) < \textsc{depth}(u)$.
Assume first that $\textsc{depth}(v) < \textsc{depth}(u)$; see
Fig.~\ref{fig:proof:below}.
Let $x$ be the successor of $v$ on $P$.
It is clear that either $\textsc{lowpt}(v,x) < \textsc{depth}(w) =
\textsc{lowpt}(v,w)$ or $\textsc{lowpt}(v,x) = \textsc{depth}(w)$ and
$\textsc{angle}(v,x) \geq \textsc{angle}(u,w) > \textsc{angle}(v,w)$.
Therefore we traverse $(v,w)$ before $(v,x)$ and consequently also $(u,w)$
during the second DFS, a contradiction to the assumption that we discover $t$
when we traverse $(v,w)$.
The other case $\textsc{depth}(u) < \textsc{depth}(v)$ is depicted in
Fig.~\ref{fig:proof:between}.
Let $x$ be the successor of $u$ on $P$.
In this case we have
$\textsc{lowpt}(u,x) = \textsc{depth}(w) = \textsc{lowpt}(u,w)$ and
$\textsc{angle}(u,x) < \textsc{angle}(u,w)$ because otherwise an edge would
cross $C(u,w)$.
Therefore we traverse $(u,w)$ after $(u,x)$ and consequently also $(v,w)$
during the second DFS, again a contradiction to the assumption that we discover
$t$ when we traverse $(v,w)$.

\paragraph{All edges contained in the interior of $t$ are traversed
before~$(v,w)$:}
Let $e$ be an edge in the interior of $t$.
If $e$ lies on $P$, then $e$ is certainly traversed before~$(v,w)$; recall that $(v,w)$ is the last edge of $t$ that is traversed and thus, $u$ and all edges on $P$ have to be traversed before traversing $(v,w)$.
Otherwise the edge $e$ lies in the interior of at least one of the fundamental cycles
$C(b)$, $b \in \{(v,w),(u,w),(v,u),(u,v)\}$~--~where the latter three only have
to be considered if they are back edges.
Let $x$ be the deepest
ancestor of the tail of $e$ on $P$ and let $z$ be the successor of $x$ on
$C(b)$.
If $x$ is the tail of $e$, let $e = (x,y)$.
Otherwise let $y$ be the child of $x$ on the tree path to the head of $e$.
See Fig.~\ref{fig:proof:inside}.
Since $(x,y)$ is in the interior of the fundamental cycle $C(b)$, it follows
that the outermost return edge of $(x,y)$ is also in the interior of $C(b)$.
Thus $(x,y)$ must have a greater lowpoint than $b$ or an equal lowpoint
but a smaller angle.
Since $(x,z)$ is on $C(b)$, it follows that the outermost return edge of $(x,z)$
is $b$ or it has a lower lowpoint than $b$ or an equal lowpoint
but a greater or equal angle.
Hence $(x,y)$ must be traversed before $(x,z)$ in the second DFS.
This implies that $e$ was traversed before $b$.
Since $(v,w)$ was the last edge traversed among $(v,w)$, $(u,w)$, and $\{u,v\}$,
it follows that $e$ must be traversed before $(v,w)$.

\paragraph{The order of the triangles is correct:}
If a separating triangle $t'$ lies inside $t$, then we either
traverse all its edges before $(v,w)$ or we traverse two of its edges before
$(v,w)$ and the third is $(v,w)$.
In the former case, we already now that our algorithm orders $t'$ before $t$.
In the latter, let $u'$ be the third vertex of $t'$, and w.l.o.g., let $(v,w)$
be a left back edge.
$u'$ lies inside $t$, thus the size $\|\angle vwu'\|$ of the internal angle
of~$t'$ is smaller than the angle $\|\angle vwu\|$ of $t$.
We compare $t$ and $t'$ by the sizes of their internal angles and order $t'$
before $t$.

\paragraph{Linear Running Time:}
During the first DFS, we do not list the candidates for the outermost return
edge of a tree edge $(v,w)$, but instead we select it from among the outermost
return edges of the outgoing edges of $w$ when popping $w$ from the DFS stack.
We use LSD-radix sort on two-digit numbers
\begin{inparaenum}[(i)]
    \item
        to sort all edges together before the second DFS (first by ascending
        angle, then stably by descending lowpoint) and
    \item
        to sort the separating triangles after the second DFS has been completed
        (first by the size of the internal angle, then stably by the point in time the
        triangle was discovered during the second DFS).
\end{inparaenum}
After sorting all edges together before the
second DFS, we partition the sorted list by the tail of the edges in order to
obtain the ordered lists of outgoing edges per vertex.
This way both of the sorting steps run in linear time.
\end{proof}

\bibliography{4connected}

\clearpage

\appendix

\section*{Appendix: Pseudocode}\label{APP}

In this appendix we provide pseudocode for Step~2 of our algorithm, i.e., for ordering the separating triangles from innermost to outermost. Observe that global variables are in small caps, local variables are single letters set in italics; uninitialized values are considered to be set to $\bot$.

\medskip
\small

\begin{algorithm}[H]
    \caption{Order separating triangles}
    \DontPrintSemicolon
    \KwIn{Graph $G = (V,E)$, list $T$ of separating triangles, vertex $r$ on outer face}
    \KwOut{Order \textsc{triangleOrder} of separating triangles}
    $\textsc{depth}[r] \gets 0$\;
    $\textsc{parentEdge}[r] \gets$ virtual edge in outer face\;
    DFS1($r$)\;
    SortEdges\;
    $\textsc{now} \gets 0$\;
    DFS2($r$)\;
    SortTriangles\;
\end{algorithm}

\bigskip

\begin{procedure}[H]
    \caption{DFS1(vertex $v$)}
    \DontPrintSemicolon
    $e \gets \textsc{parentEdge}[v]$\;
    \While{there exists some non-oriented $\{v,w\} \in E$}{
        orient $\{v,w\}$ as $(v,w)$\;
        \uIf(\tcp*[h]{tree edge}){\rm$\textsc{depth}[w] = \bot$}{
            $\textsc{parentEdge}[w] \gets (v,w)$\;
            $\textsc{depth}[w] \gets \textsc{depth}[v] + 1$\;
            $\textsc{activeChild}[v] \gets w$\;
            \underline{DFS1}($w$)\;
        }
        \Else(\tcp*[h]{back edge}){
            $(p,w) \gets \textsc{parentEdge}[w]$\;
            $c \gets \textsc{activeChild}[w]$ \tcp*[l]{successor of w on C(v,w)}
            $\textsc{lowpt}[(v,w)] \gets \textsc{depth}[w]$\;
            \uIf{$\|\angle vwc\| < \|\angle pwc\|$}{
                $(v,w)$ is a left back edge\;
                $\textsc{angle}[(v,w)] \gets \|\angle vwc\|$\;
            }
            \Else{
                $(v,w)$ is a right back edge\;
                $\textsc{angle}[(v,w)] \gets \|\angle cwv\|$\;
            }
            $\textsc{outermostReturnEdge}[(v,w)] \gets (v,w)$\;
        }
        \SetKwBlock{Begin}{\ \!$\blacktriangledown$ update outermost return edge of parent edge}{}
        \Begin{
            $o \gets \textsc{outermostReturnEdge}[e]$\;
            $o' \gets \textsc{outermostReturnEdge}[(v,w)]$\;
            \If{\rm$o = \bot \lor (-\textsc{lowpt}[o'], \textsc{angle}[o']) > (-\textsc{lowpt}[o], \textsc{angle}[o])$}{
                $\textsc{outermostReturnEdge}[e] \gets o'$\;
            }
        }
    }
    \SetKwBlock{Begin}{\ \!$\blacktriangledown$ parent edge inherits from its outermost return edge}{}
    \Begin{
        $\textsc{lowpt}[e] \gets \textsc{lowpt}[\textsc{outermostReturnEdge}[e]]$\;
        $\textsc{angle}[e] \gets \textsc{angle}[\textsc{outermostReturnEdge}[e]]$\;
    }
\end{procedure}

\begin{procedure}[H]
    \caption{SortEdges()}
    \DontPrintSemicolon
    $A \gets \textrm{CountingSort}(E, e \mapsto \textsc{angle}[e])$\;
    $L \gets \textrm{CountingSort}(A, e \mapsto |V| - \textsc{lowpt}[e])$\; 
    \For{$(v,w) \in L$}{
        $\textrm{append}(\textsc{edgeOrder}[v], (v,w))$\;
    }
\end{procedure}

\bigskip

\begin{procedure}[H]
    \caption{DFS2(vertex $v$)}
    \DontPrintSemicolon
    \For{\rm$(v,w) \in \textsc{edgeOrder}[v]$}{
        $\textsc{edgeTime}[(v,w)] \gets \textsc{now}$\;
        $\textsc{now} \gets \textsc{now} + 1$\;
        \If(\tcp*[h]{tree edge}){\rm$\textsc{depth}[v] + 1 = \textsc{depth}[w]$}{
            \underline{DFS2}($w$)\;
        }
    }
\end{procedure}

\bigskip

\begin{procedure}[H]
    \caption{SortTriangles()}
    \DontPrintSemicolon
    \For{$t \in T$}{
        $\displaystyle (v,w) \gets \argmax_{\textrm{edge } e \in t} \textsc{edgeTime}[e]$\;
        $u \gets$ third vertex on $t$\;
        $\textsc{time}[t] \gets \textsc{edgeTime}[(v,w)]$\;
        \uIf{$(v,w)$ is a left back edge}{
            $\textsc{referenceEdge}[t] \gets (w,v)$\;
            $\textsc{internalAngle}[t] \gets \|\angle vwu\|$\;
        }
        \Else{
            $\textsc{referenceEdge}[t] \gets (v,w)$\;
            $\textsc{internalAngle}[t] \gets \|\angle uwv\|$\;
        }
    }
    $A \gets \textrm{CountingSort}(T, t \mapsto \textsc{internalAngle}[t])$\;
    $\textsc{triangleOrder} \gets \textrm{CountingSort}(A, t \mapsto \textsc{time}[t])$\;
\end{procedure}

\end{document}